# Eigen electric moments of magnetic-dipolar modes in quasi-2D ferrite disk particles

M. Sigalov, E.O. Kamenetskii, and R. Shavit

Ben-Gurion University of the Negev, Beer Sheva 84105, Israel

August 20, 2007

**Abstract**

A property associated with a vortex structure becomes evident from an analysis of confinement phenomena of magnetic oscillations in a quasi-2D ferrite disk with a dominating role of magnetic-dipolar (non-exchange-interaction) spectra. The vortices are guaranteed by the chiral edge states of magnetic-dipolar modes which result in appearance of eigen electric moments oriented normally to the disk plane. Due to the eigen-electric-moment properties, a ferrite disk placed in a microwave cavity is strongly affected by the cavity RF electric field with a clear evidence for multi-resonance oscillations. For different cavity parameters, one may observe the "resonance absorption" and "resonance repulsion" behaviors.

In a dot of a ferromagnetic material of micrometer or submicrometer size, a curling spin configuration – that is, a magnetization vortex – has been proposed. The vortex consists of an in-plane, flux-closure magnetization distribution and a central core whose magnetization is perpendicular to the dot plane. It has been shown that under certain conditions a vortex structure will be stable because of competition between the exchange and dipole interactions [1]. In spite of the fact that vortices can appear in different kinds of physical phenomena, yet such "swirling" entities seem to elude an all-inclusive definition. It appears that the character of magnetic vortices in magnetically soft "small" (with the dipolar and exchange energy competition) and magnetically saturated "big" (when the exchange is neglected) ferrite disks is very different. A magnetization vortex in a magnetically soft sample cannot be characterized by some invariant, such as the flux of vorticity. So a vorticity thread may not be defined for the magnetization vortex [2]. At the same time, in magnetically saturated samples with magnetic-dipolar vortices one can observe the flux of the pseudo-electric (gauge) fields [3, 4]. The vortices of magnetic-dipolar-mode (MDM) oscillations in a ferrite disk become apparent due to the symmetry breaking effects which result in appearance of eigen electric moments oriented normally to the disk plane [3 – 5].

   A property associated with a vortex structure in a ferrite disk with a dominating role of magnetic-dipolar (non-exchange-interaction) spectra becomes evident from an analysis of confinement phenomena of magnetic-dipolar oscillations. It has been shown [3, 4, 6] that for MDMs in a ferrite disk one has evident quantum-like attributes. The spectrum is characterized by energy eigenstate oscillations. It appears, however, that because of the boundary condition on a lateral surface of a ferrite disk, MS-potential eigen functions cannot be considered as single-valued functions. This fact raises a question about validity of the energy orthogonality relation for the MDMs. The most basic implication of the existence of a phase factor in eigen functions is operative in the case on the border ring region. It follows that in order to cancel the "edge anomaly", the boundary excitation must be described by chiral states [3, 4].

The topological effects in the MDM ferrite disk are manifested through the generation of relative phases which accumulate on the boundary wave functions $\delta_\pm$ [3, 4]:

$$\delta_\pm \equiv f_\pm e^{-iq_\pm \theta}. \tag{1}$$

The quantities $q_\pm$ are equal to $\pm l\frac{1}{2}$, $l = 1, 3, 5, ...$ For amplitudes $f$ we have $f_+ = -f_-$ with normalization $|f_\pm| = 1$. To preserve the single-valued nature of the membrane functions of the MDM oscillations, functions $\delta_\pm$ must change its sign when a disk angle coordinate $\theta$ is rotated by $2\pi$ so that $e^{-iq_\pm 2\pi} = -1$. A sign of a full chiral rotation, $q_+\theta = \pi$ or $q_-\theta = -\pi$, should be correlated with a sign of the parameter $i\mu_a$ – the off-diagonal component of the permeability tensor $\vec{\mu}$. This becomes evident from the fact that a sign of $i\mu_a$ is related to a precession direction of a magnetic moment $\vec{m}$. In a ferromagnetic resonance, the bias field sets up a preferential precession direction. It means that for a normally magnetized ferrite disk with a given direction of a DC bias magnetic field, there are two types of resonant oscillations, which we conventionally designate as the (+) resonance and the (−) resonance. For the (+) resonance, a direction of an edge chiral rotation coincides with the precession magnetization direction, while for the (−) resonance, a direction of an edge chiral rotation is opposite to the precession magnetization direction. Fig. 1 gives an example of the edge-function chiral rotation in correlation with the RF magnetization evolution for the (+) resonance.

For a ferrite disk with $r$ and $\theta$ in-plane coordinates and normal-axis $z$ coordinate, the total MS-potential function $\psi$ is represented as a product of three functions:

$$\psi = \tilde{\eta}(r,\theta)\,\xi(z)\,\delta_\pm, \tag{2}$$

where $\tilde{\eta}(r,\theta)$ is a single-valued membrane function, $\xi(z)$ is the function characterizing $z$-distribution of the MS potential in a ferrite disk, and $\delta_\pm$ is a double-valued edge (spin-coordinate-like) function. We may introduce a "spin variable" $\sigma$, representing the orientation of the "spin moment" and two doubled-valued wave functions, $\delta_+(\sigma)$ and $\delta_-(\sigma)$, the former corresponding to the eigen value $q_+ = +\frac{1}{2}$ and the latter to the eigen value $q_- = -\frac{1}{2}$. The two wave functions are normalized and mutually orthogonal, so that they satisfy the equations $\int \delta_+^2(\sigma)\,d\sigma = 1$, $\int \delta_-^2(\sigma)\,d\sigma = 1$, and $\int \delta_+(\sigma)\delta_-(\sigma)\,d\sigma = 0$. A wave function $\psi$ is then a function of four coordinates, three positional coordinates such as $r, \theta$, and $z$, and the "spin coordinate" $\sigma$. We write $\psi = \tilde{\eta}(r,\theta)\,\xi(z)\,\delta_+(\sigma)$ and $\psi = \tilde{\eta}(r,\theta)\,\xi(z)\,\delta_-(\sigma)$ as two wave functions corresponding to the positional wave function $\tilde{\eta}(r,\theta)\,\xi(z)$, which is a solution of the Walker equation for a ferrite disk with the so-called essential boundary conditions [6].

It is evident that for a ferrite disk of radius $\mathfrak{R}$, circulation of gradient

$$\vec{\nabla}_\theta \delta_\pm = \frac{1}{\mathfrak{R}}\frac{\partial \delta_\pm}{\partial \theta}\bigg|_{r=\mathfrak{R}} \vec{e}_\theta = -i\frac{q_\pm f_\pm}{\mathfrak{R}} e^{-iq_\pm \theta}\vec{e}_\theta \tag{3}$$

along a disk border contour $C = 2\pi\Re$ gives a nonzero quantity when $q_\pm$ is a number divisible by $\frac{1}{2}$. We consider the quantity $\nabla_\theta \delta_\pm$ as the velocity of an irrotational "border" flow:

$$\left(\vec{v}_\theta\right)_\pm \equiv \vec{\nabla}_\theta \delta_\pm .\qquad(4)$$

In such a sense, functions $\delta_\pm$ are the velocity potentials. Circulation of $\left(\vec{v}_\theta\right)_\pm$ along a contour $C$ is equal to $\oint_C \left(\vec{v}_\theta\right)_\pm \cdot d\vec{C} = \Re \int_0^{2\pi} \nabla_\theta \delta_\pm \, d\theta = -2 f_\pm$.

In a case of a cylindrical ferrite disk, a single-valued membrane function is represented as $\tilde{\eta}(r,\theta) = R(r)\phi(\theta)$, where $R(r)$ is described by the Bessel functions and $\phi(\theta) \sim e^{-i\nu\theta}$, $\nu = \pm 1, \pm 2, \pm 3....$ Taking into account the "orbital" function $\phi(\theta)$, we may consider the quantity $\left[\vec{\nabla}_\theta (\phi\, \delta_\pm)\right]_{r=\Re}$ as the total ("orbital" and "spin") velocity of an irrotational "border" flow:

$$\left(\vec{V}_\theta\right)_\pm \equiv \left[\vec{\nabla}_\theta (\phi\, \delta_\pm)\right]_{r=\Re} .\qquad(5)$$

It is evident that

$$\left(\vec{V}_\theta\right)_\pm = -i \frac{(\nu + q_\pm) f_\pm}{\Re} e^{-i(\nu+q_\pm)\theta} \vec{e}_\theta .\qquad(6)$$

For a given membrane function $\tilde{\eta}$ and given $z$-distribution of the MS potential, $\xi(z)$, we can define now the strength of a vortex of a whole disk as

$$s_\pm^e \equiv R_{r=\Re} \int_0^d \xi(z) dz \oint_C \left(\vec{V}_\theta\right)_\pm \cdot d\vec{C} = \Re R_{r=\Re} \int_0^d \xi(z) dz \int_0^{2\pi} \left(\vec{V}_\theta\right)_\pm \cdot \vec{e}_\theta \, d\theta = -2 f_\pm R_{r=\Re} \int_0^d \xi(z) dz ,\qquad(7)$$

where $d$ is a disk thickness.

The quantity $\left(\vec{V}_\theta\right)_\pm$ has a clear physical meaning. In the spectral problem for MDM ferrite disks, non-singlevaluedness of the MS-potential wave function appears due to the border term which is defined as $-i\mu_a (H_\theta)_{r=\Re}$. This border term arises from the demand of conservation of the magnetic flux density [3, 4]. It is evident that an annual magnetic field on the border circle, $(H_\theta)_{r=\Re}$, is expressed as

$$\left(\left(\vec{H}_\theta(z)\right)_\pm\right)_{r=\Re} = -\xi(z)\left(\vec{V}_\theta\right)_\pm .\qquad(8)$$

We define now an angular moment $\vec{a}_\pm^e$:

$$a_\pm^e \equiv \int_0^d \oint_C \left[-i\mu_a (H_\theta)_{r=\Re}\right] \vec{e}_\theta \cdot d\vec{C} \, dz = i\mu_a \, s_\pm^e .\qquad(9)$$

This angular moment can be formally represented as a result of a circulation of a quantity, which we call a density of an effective boundary magnetic current $\vec{i}^{\,m}$:

$$a_\pm^e = 4\pi \int_0^d \oint_C \vec{i}_\pm^{\,m} \cdot d\vec{C}\, dz, \tag{10}$$

where $\vec{i}_\pm^{\,m} \equiv \rho^m \left(\vec{V}_\theta\right)_\pm$ and $\rho^m \equiv i\dfrac{\mu_a}{4\pi}\xi\, R_{r=\Re}$.

In our continuous-medium model, a character of the magnetization motion becomes apparent via the gyration parameter $\mu_a$ in the boundary term for the spectral problem. There is magnetization motion through a non-simply-connected region. On the edge region, the chiral symmetry of the magnetization precession is broken to form a flux-closure structure. The edge magnetic currents can be observable only via its circulation integrals, not pointwise. This results in the moment oriented along a disk normal. It was shown experimentally [5] that such a moment has a response in an external RF electric field. This clarifies a physical meaning of a superscript "e" in designations of $s_\pm^e$ and $a_\pm^e$. In a ferrite disk particle, the vector $\vec{a}^e$ is an electric moment characterizing by special symmetry properties. There are the anapole-moment properties [3, 4].

The purpose of this letter is to analyze the spectral distribution of eigen electric (anapole) moments of a MDM ferrite disk. Our experimental studies of absorption spectra for a ferrite disk placed in a microwave cavity are aimed to investigate possible mechanisms of interaction of the disk eigen electric moments with an external RF electric field and to verify the proposed theoretical model for the anapole moment oscillations.

Experimental results showed in [5] leaved unclear the physics of a possible mechanism of interaction between a MDM ferrite disk and external RF electric fields. One can suppose a classical mechanism of interaction between the disk eigen electric moment and the cavity $E$-field. But the question is why this interaction gives the multiresonance spectral pictures similar to the pictures shown in well known experiments with a ferrite disk placed in an external RF magnetic field [7]. The main aspects concern the question how one may obtain effective resonance interactions between the edge double valued functions (determining the anapole-moment properties) and the membrane single valued functions (determining the energy eigen states). To understand this mechanism we suggest here the following qualitative model. Fig. 2 (a) shows the double-valued functions $\delta_+(\theta')$ and $\nabla_{\theta'}\delta_+(\theta')$ for $q = +\dfrac{1}{2}$. Here we use designation $\theta'$ to distinguish a "spin" angular coordinate from a regular angle coordinate. Because of the edge-function chiral rotation [3, 4], one has to select only positive derivatives: $\dfrac{\partial \delta_+}{\partial \theta'} > 0$. The corresponding parts of the graphs are distinguished in Fig. 2 (a) by bold lines. Figs. 2 (b) and 2 (c) give two cases of the membrane single valued functions $\tilde{\eta}(\theta)$ which may lead to resonance "spin-orbital" interactions. It is evident that in a case of Fig. 2 (b), a positive half of function $\delta_+(\theta')$ is phased for a resonance interaction with the positive halves of function $\tilde{\eta}(\theta)$, while in a case of Fig. 2 (c) a positive half of function $\delta_+(\theta')$ is phased for a resonance interaction with the negative halves of function $\tilde{\eta}(\theta)$. This results in a non-zero integral in Eq. (7) and explains how the double-valued-function spins precessing on the edge region may interact with the single-valued-function spins precessing in the core region. For a case of Fig. 2 (b) one has the "resonance absorption" and for a case of Fig. 2 (c) one has the "resonance repulsion". Both types of interactions are equiprobable. One may expect that for different cavity parameters, two the above cases, the "resonance absorption" and "resonance repulsion", can be exhibited separately. One may also expect that in a certain situation transitions between these two resonance behaviors can be demonstrated. Our experiments clearly verify this model of resonance interactions.

In experiments, we used a disk sample of a diameter $2\Re = 3 \, \text{mm}$ made of the YIG film on the GGG substrate (the YIG film thickness $d = 50 \, \text{mkm}$, saturation magnetization $4\pi M_0 = 1880 \, \text{G}$, linewidth $\Delta H = 0.8 \, \text{Oe}$) and a short-wall rectangular-waveguide cavity with an entering iris. A normally magnetized ferrite disk was placed in a cavity in a maximal RF electric field of the $TE_{102}$ mode and was oriented normally to the *E*-field (Fig. 3). In Fig. 4, showing the frequency dependence of the cavity reflection coefficient (CRC), we point out three characteristic frequencies used in experiments.

An analysis made in [5] allows clearly specify the main spectral features of an interaction of a MDM ferrite disk with a cavity RF electric field. Fig 5 (a) shows a typical multiresonance spectrum of such an interaction. This is a dependence of the absolute value of the CRC on a bias magnetic field obtained for a critically coupled cavity at the resonance frequency $f_1 = 7.085$ GHz. The digits are the MDM numbers. For the cavities with reduced *Q*-factors and resonant at the same frequency (to preserve the resonance frequency we used small tuning elements), the character of the ferrite-disk spectrum remains the same [see Figs. 5 (b), (c)]. Following the above analysis one can conclude that the spectra in Figs. 5 (a), (b) and (c) correspond to the "resonance repulsion". When we put a small piece of a metallic wire (made of copper) above a ferrite disk and parallel to the cavity *E*-field, we obtained a very strong interaction between the disk and cavity [Fig. 5 (d)]. In this case we can discern two fundamental aspects. First of all, the fact that an additional small capacitive coupling strongly affects on magnetic oscillation proves, once again, the presence of the electric-dipole moments of the MDMs in a quasi-2D ferrite disk. Secondly, we see very unique features in the spectral picture. There are sharp jumps of the CRCs in the regions of the disk resonance peaks. It can be definitely supposed that these jumps are caused by sharp phase transitions between two $2\pi$-behaviors shown in Figs. 2 (b) and (c).

To investigate more in details transitions between behaviors of the "resonance repulsion" and "resonance absorption" we analyzed the ferrite disk spectra measured at different frequencies. These frequencies, $f_1, f_2,$ and $f_3$, correspond to different positions on the resonance curve of the cavity (see Fig. 4). The multiresonance spectral pictures for these frequencies are shown in Fig 6. Since the permeability tensor parameters are dependent both on frequency and a bias magnetic field, we were able to match the peak position by small variations of a bias field. The fields corresponding to the first peaks are adduced in the figure.

The spectrum in Fig. 6 (a), corresponding to $f_1$ [and being the same as the spectrum in Fig. 5 (b)], represents the "resonance repulsion" behavior. At the same time, the spectrum in Fig. 6 (c), corresponding to $f_3$, clearly demonstrates the "resonance absorption" behavior. It becomes evident that the spectrum in Fig. 6 (b), corresponding to $f_2$, shows the transitions between the "resonance repulsion" and "resonance absorption". A qualitative explanation of the observed three cases could be the following. Since at frequency $f_1$ the cavity is "viewed" by the incoming signal as an active load, one can clearly observe the "resonance repulsion" due to a ferrite disk. Contrary, at frequency $f_3$ the cavity is characterized mainly as a reactive load. In this case one observes the "resonance absorption" behavior. At frequency $f_2$ both cases are mixed and a transitional behavior takes place. It is worth noting that for transitional behaviors shown in Figs. 5 (d) and 6 (b), the "between-peak derivatives" of CRCs with respect to the bias field are of different signs.

For the above disk parameters used in experiments we calculated amplitudes of eigen electric moments of MDMs based on Eq. (10). The calculations of functions $\xi(z)$ and $R(r)$ were made for "orbital" azimuth number $\nu = 1$ and for the essential boundary conditions based on the methods used in [6]. The calculation results of the eigen-electric-moment amplitudes are shown in Fig. 7. To compare the calculation results with the experimental ones we took the measured relative peak amplitudes. We "tied" together the calculated and measured amplitudes of the first-mode peaks and normalized them to unit. Evidently that since the mode peak amplitudes were measured with respect to a bias magnetic field at the constant frequency, we had negligibly small "from-mode-to-mode" variations of the cavity *E*-field amplitudes. So the measured relative peak amplitudes should correspond to experimental MDM distributions of the eigen electric moments. Fig. 7 shows relatively good agreement between the experimental and calculation results. Some disagreement can be explained by certain inaccuracy in precise experimental characterization of amplitudes of very sharp resonant peaks.

In conclusion, we have to note that the eigen electric moments of a ferrite disk arises not from the classical curl electric fields of magnetostatic oscillations. At the same time, any induced electric polarization effects in YIG or GGG materials are beyond the frames of the observed multiresonance spectra. We sum up that in this letter we calculated the spectral distribution of the eigen electric (anapole) moments of a MDM ferrite disk. We discussed a model which gives a possible picture of interaction of the MDM oscillations with external RF electric fields. Our experimental results, shown in this letter, convincingly confirmed the proposed model of anapole moment oscillations caused by edge chiral rotations in a MDM ferrite disk. We demonstrated good correlation between the calculated and experimental results.

**Figure captions**
Fig. 1. Edge-function chiral rotation in correlation with the RF magnetization evolution for the (+) resonance.
Fig. 2. The model explaining resonance interactions of the MDM ferrite disk with the cavity electric field.
Fig. 3. A waveguide cavity with an enclosed ferrite disk.
Fig. 4. Frequency dependence of the cavity reflection coefficient.
Fig. 5. Spectral pictures of reflection coefficients for different cavity structures: (a) Critically coupled cavity; (b) Non critically coupled cavity; (c) Cavity with an inserted loss material; (d) Critically coupled cavity with inserted small metallic wire above a ferrite disk.
Fig. 6. Spectral pictures of reflection coefficients obtained at different frequencies.
Fig. 7. Calculated and measured electric moment amplitudes versus MDM numbers.

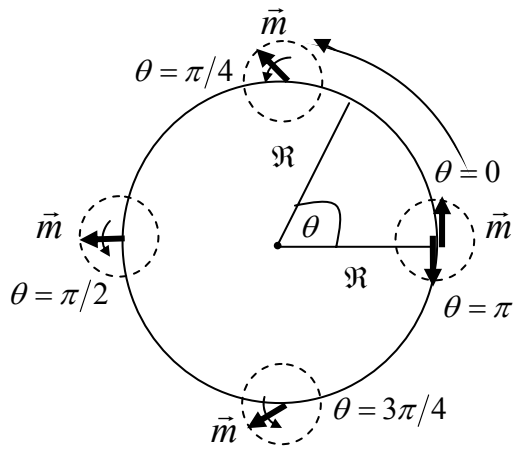

Fig. 1. Edge-function chiral rotation in correlation with the RF magnetization evolution for the (+) resonance.

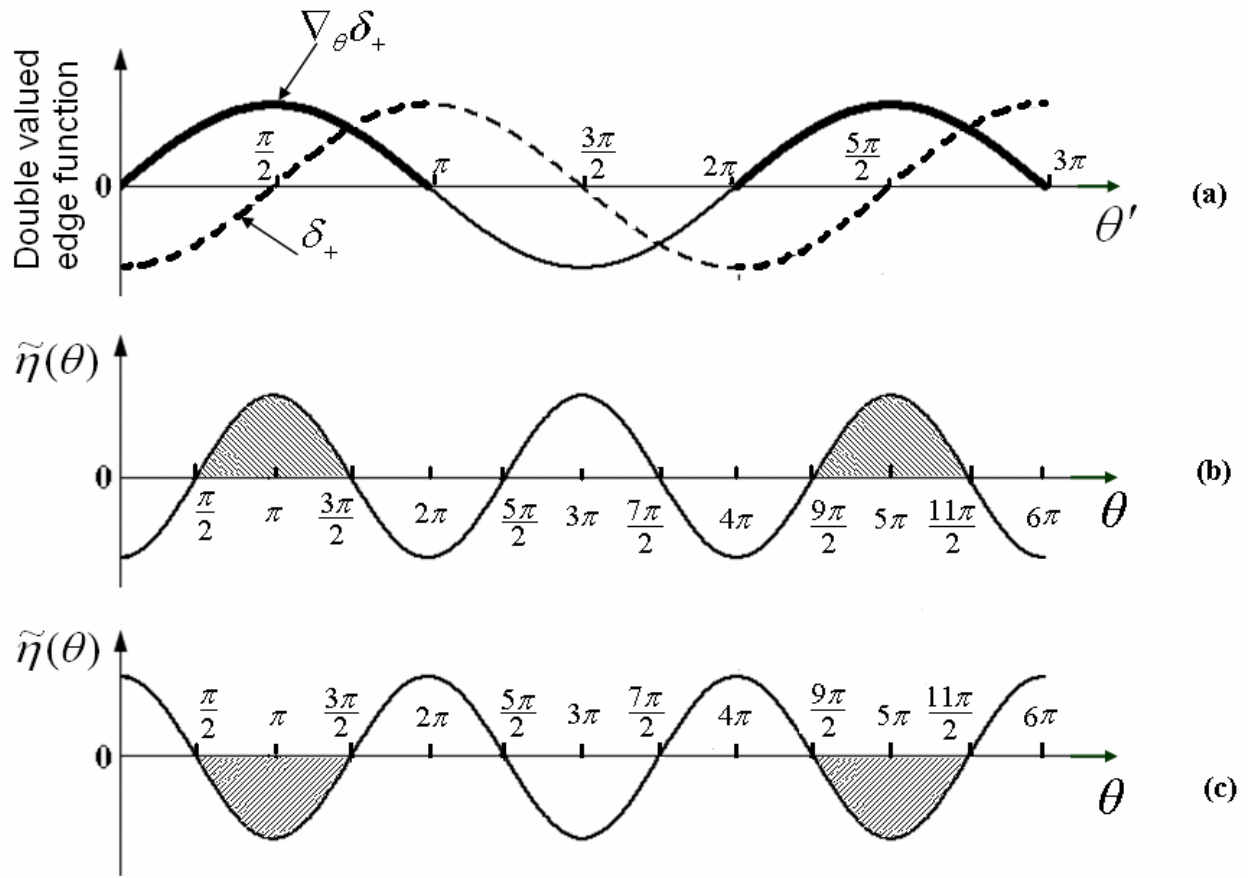

Fig. 2. The model explaining resonance interactions of the MDM ferrite disk with the cavity electric field.

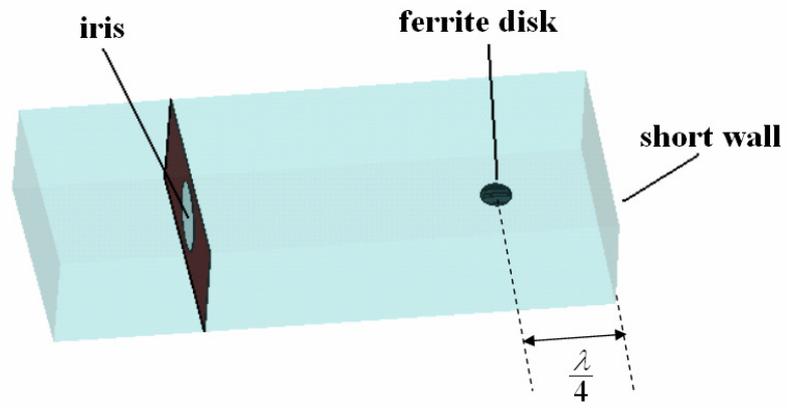

(b)

Fig. 3. A waveguide cavity with an enclosed ferrite disk.

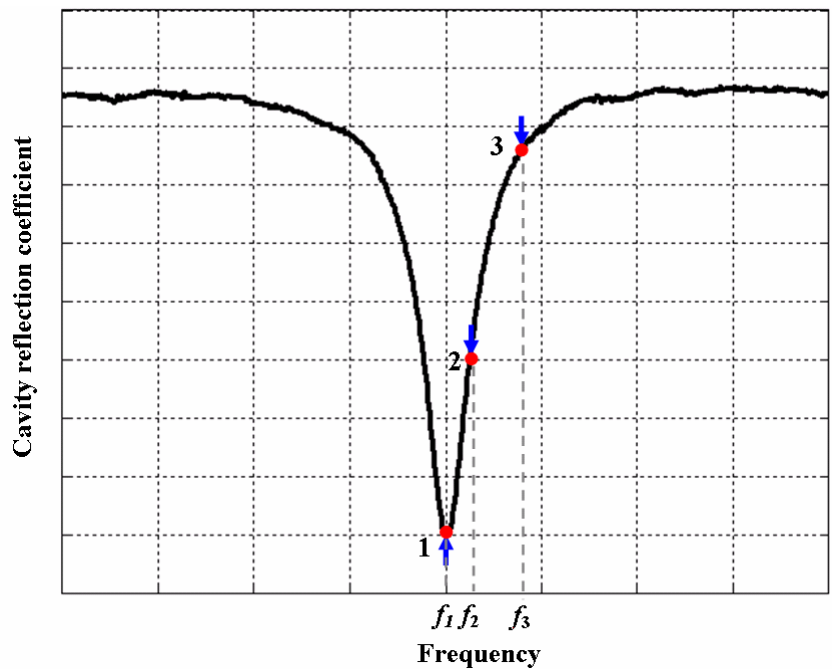

Fig. 4. Frequency dependence of the cavity reflection coefficient.

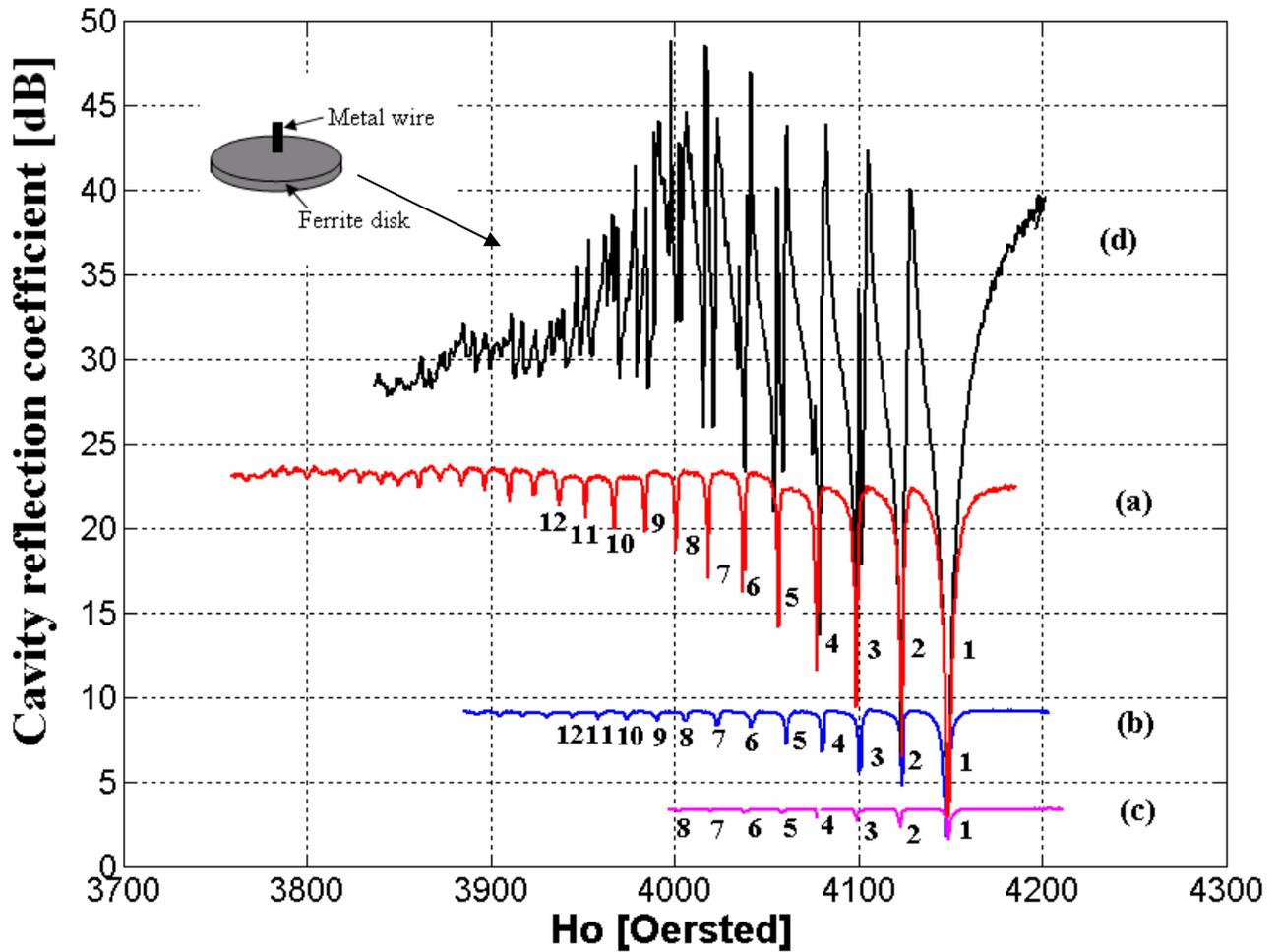

Fig. 5. Spectral pictures of reflection coefficients for different cavity structures: (a) Critically coupled cavity; (b) Non critically coupled cavity; (c) Cavity with an inserted loss material; (d) Critically coupled cavity with inserted small metallic wire above a ferrite disk.

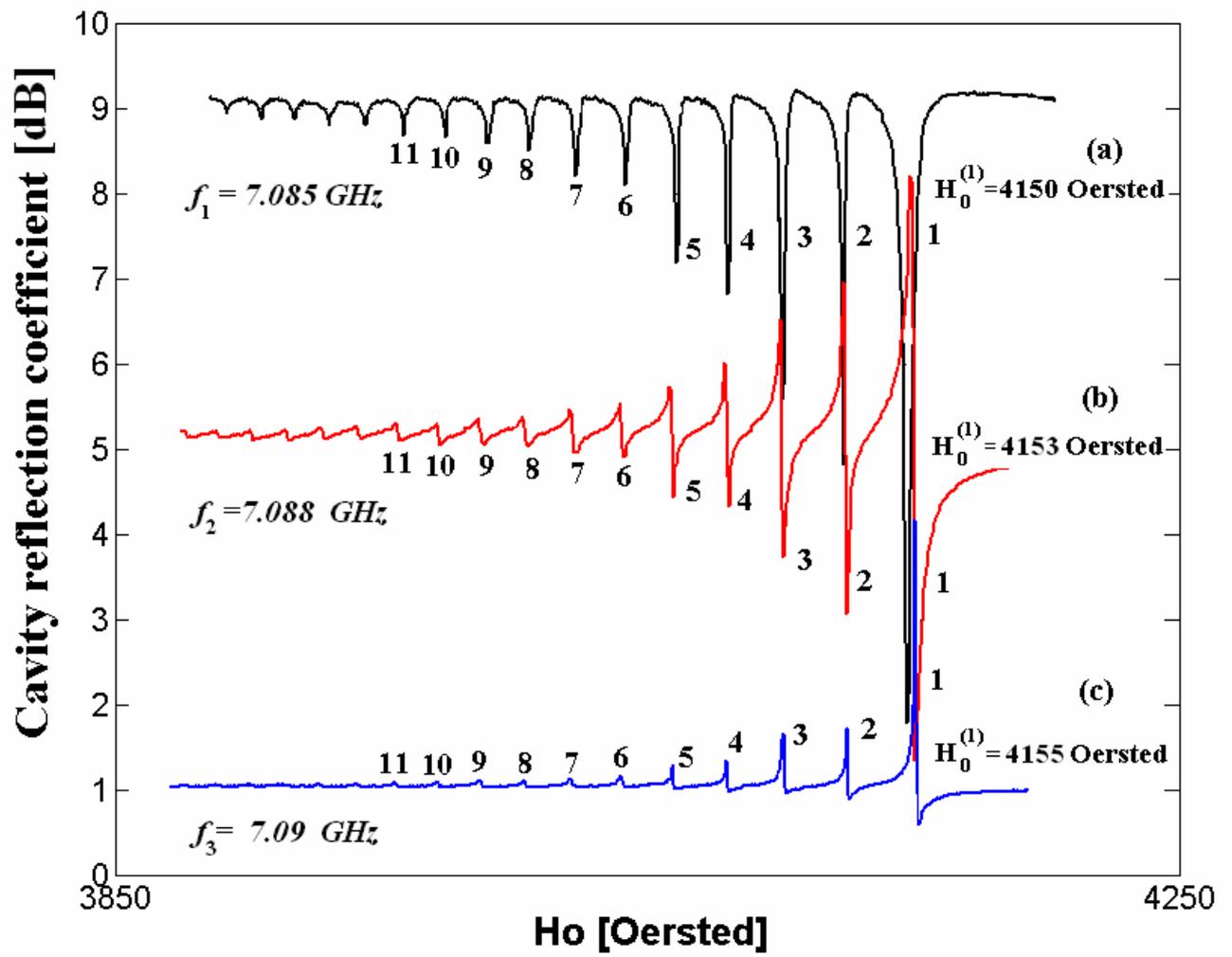

Fig. 6. Spectral pictures of reflection coefficients obtained at different frequencies.

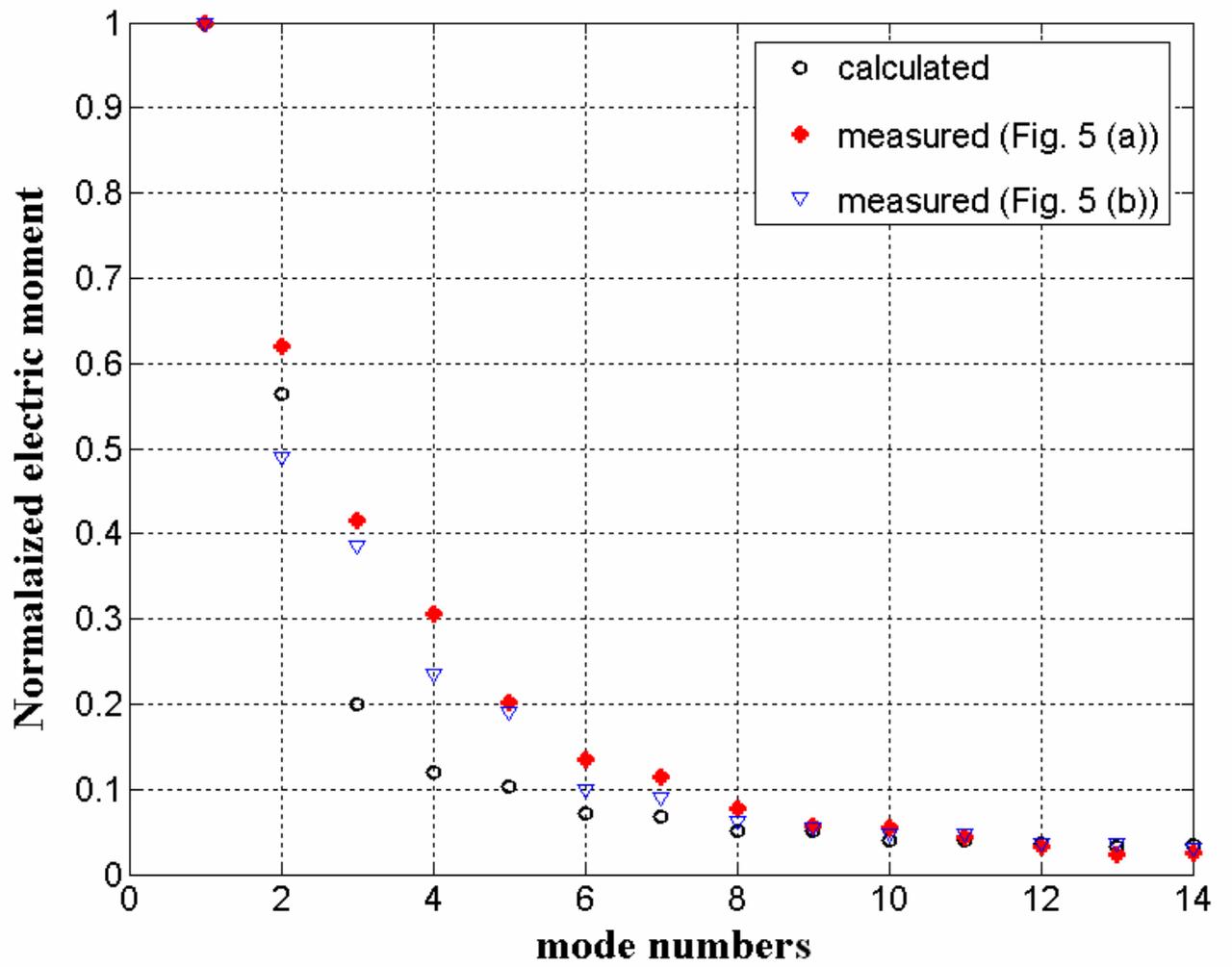

Fig. 7. Calculated and measured electric moment amplitudes versus MDM numbers.